\documentstyle[epsf]{mn}

\newif\ifAMStwofonts

\ifoldfss
  \ifCUPmtlplainloaded \else
    \NewTextAlphabet{textbfit} {cmbxti10} {}
    \NewTextAlphabet{textbfss} {cmssbx10} {}
    \NewMathAlphabet{mathbfit} {cmbxti10} {} 
    \NewMathAlphabet{mathbfss} {cmssbx10} {} 
  \fi
  \ifAMStwofonts
    \ifCUPmtlplainloaded \else
      \NewSymbolFont{upmath} {eurm10}
      \NewSymbolFont{AMSa} {msam10}
      \NewMathSymbol{\upi}     {0}{upmath}{19}
      \NewMathSymbol{\umu}     {0}{upmath}{16}
      \NewMathSymbol{\upartial}{0}{upmath}{40}
      \NewMathSymbol{\leqslant}{3}{AMSa}{36}
      \NewMathSymbol{\geqslant}{3}{AMSa}{3E}

       \let\le=\leqslant
       \let\ge=\geqslant
    \fi
  \fi
\fi 

\ifnfssone
  \newmathalphabet{\mathit}
  \addtoversion{normal}{\mathit}{cmr}{m}{it}
  \addtoversion{bold}{\mathit}{cmr}{bx}{it}
  \newmathalphabet{\mathbfit} 
  \addtoversion{normal}{\mathbfit}{cmr}{bx}{it}
  \addtoversion{bold}{\mathbfit}{cmr}{bx}{it}
  \newmathalphabet{\mathbfss} 
  \addtoversion{normal}{\mathbfss}{cmss}{bx}{n}
  \addtoversion{bold}{\mathbfss}{cmss}{bx}{n}
  \ifAMStwofonts
    \ifCUPmtlplainloaded \else
      \UseAMStwoboldmath
      \makeatletter
      \new@mathgroup\upmath@group
      \define@mathgroup\mv@normal\upmath@group{eur}{m}{n}
      \define@mathgroup\mv@bold\upmath@group{eur}{b}{n}
      \edef\UPM{\hexnumber\upmath@group}
      \new@mathgroup\amsa@group
      \define@mathgroup\mv@normal\amsa@group{msa}{m}{n}
      \define@mathgroup\mv@bold\amsa@group{msa}{m}{n}
      \edef\AMSa{\hexnumber\amsa@group}
      \makeatother
      \mathchardef\upi="0\UPM19
      \mathchardef\umu="0\UPM16
      \mathchardef\upartial="0\UPM40
      \mathchardef\leqslant="3\AMSa36
      \mathchardef\geqslant="3\AMSa3E

       \let\le=\leqslant
       \let\ge=\geqslant
    \fi
  \fi
\fi 

\ifnfsstwo
  \DeclareMathAlphabet{\mathbfit}{OT1}{cmr}{bx}{it}
  \SetMathAlphabet\mathbfit{bold}{OT1}{cmr}{bx}{it}
  \DeclareMathAlphabet{\mathbfss}{OT1}{cmss}{bx}{n}
  \SetMathAlphabet\mathbfss{bold}{OT1}{cmss}{bx}{n}
  \ifAMStwofonts
    \ifCUPmtlplainloaded \else
      \DeclareSymbolFont{UPM}{U}{eur}{m}{n}
      \SetSymbolFont{UPM}{bold}{U}{eur}{b}{n}
      \DeclareSymbolFont{AMSa}{U}{msa}{m}{n}
      \DeclareMathSymbol{\upi}{0}{UPM}{"19}
      \DeclareMathSymbol{\umu}{0}{UPM}{"16}
      \DeclareMathSymbol{\upartial}{0}{UPM}{"40}
      \DeclareMathSymbol{\leqslant}{3}{AMSa}{"36}
      \DeclareMathSymbol{\geqslant}{3}{AMSa}{"3E}

       \let\le=\leqslant
       \let\ge=\geqslant
    \fi
  \fi
\fi 

\ifCUPmtlplainloaded \else
  \ifAMStwofonts \else 
    \def\upi{\pi}
    \def\umu{\mu}
    \def\upartial{\partial}
  \fi
\fi

\title[Time-dependent quasi-spherical accretion]{Time-dependent
quasi-spherical accretion}
\author[G. I. Ogilvie]{G. I. Ogilvie\\
Max-Planck-Institut f\"ur Astrophysik, Karl-Schwarzschild-Stra\ss e 1,
Postfach 1523, D-85740 Garching bei M\"unchen, Germany}
\pagerange{\pageref{firstpage}--\pageref{lastpage}}
\pubyear{1998}

\begin{document}

\maketitle

\label{firstpage}

\begin{abstract}
Differentially rotating, `advection-dominated' accretion flows are
considered in which the heat generated by viscous dissipation is
retained in the fluid.  The equations of time-dependent
quasi-spherical accretion are solved in a simplified one-dimensional
model that neglects the latitudinal dependence of the flow.  A
self-similar solution is presented that has finite size, mass, angular
momentum and energy.  This may be expected to be an attractor for the
initial-value problem in which a cool and narrow ring of fluid
orbiting around a central mass heats up, spreads radially and is
accreted.  The solution provides some insight into the dynamics of
quasi-spherical accretion and avoids many of the strictures of the
steady self-similar solution of Narayan \& Yi.  Special attention is
given to the astrophysically important case in which the adiabatic
exponent $\gamma=5/3$; even in this case, the flow is found to be
differentially rotating and bound to the central object, and accretion
can occur without the need for powerful outflows.
\end{abstract}


\section{Introduction}

Accretion flows in which a significant fraction of the heat generated
by viscous dissipation is retained in the fluid, rather than radiated
away, have been the subject of considerable attention in recent years
(e.g. Narayan \& Yi 1994, 1995a,b; Abramowicz et al. 1995).  These
`advection-dominated' flows occupy an intermediate position between
the spherically symmetric accretion flow of a non-rotating fluid
(Bondi 1952) and the cool, thin disc of classical accretion-disc
theory (e.g. Pringle 1981).  Although their self-consistency may be
contingent on certain unresolved issues of plasma physics, they have
been considered relevant to a variety of astrophysical objects, in
particular to explain the low luminosity of the Galactic Centre
(e.g. Narayan et al. 1998).  A simple algebraic model describing the
fluid-dynamical aspects of such a flow, which was assumed to be
steady, axisymmetric and radially self-similar, was presented by
Narayan \& Yi (1994, hereafter NY).  They adopted a set of approximate
spherically (or vertically) averaged equations which were largely
justified by subsequent calculations that included the latitudinal
dependence (Narayan \& Yi 1995a; see also Gilham 1981).

Two properties of this solution have been emphasized and identified as
potential difficulties with its application.  First, in the
astrophysically important case in which the adiabatic exponent
$\gamma=5/3$, the solution of NY coincides with that of Bondi (1952).
This is spherically symmetric and non-rotating, has no viscous torque
or dissipation, and has been considered to depart too far from the
concept of an accretion disc.  In particular, it is not clear how such
a solution could connect to the initial or external conditions if the
supplied matter had any angular momentum.  To circumvent this
difficulty, some authors have investigated the possibility that the
effective value of $\gamma$ might be less than $5/3$ if the magnetic
field contributed a significant fraction of the total pressure
(Narayan \& Yi 1995b; Quataert \& Narayan 1998).

Secondly, the Bernoulli parameter of the flow is positive.  This has
been interpreted as meaning that the fluid is unbound and would
generate a powerful outflow if the constraints of the model were
somehow relaxed.  Xu \& Chen (1997) therefore generalized the work of
Narayan \& Yi (1995a) by constructing self-similar solutions that
involve inflow at equatorial latitudes and outflow at polar latitudes.
However, to achieve this, the scalings of physical quantities with
radius were adjusted arbitrarily from their natural values so that
only an asymptotically small fraction of the mass reaches the central
object, and in this sense the flow is hardly an accretion flow.  The
inflow is forced to turn into an outflow because the central object by
assumption accretes essentially no mass.  These solutions are
relatively unconstrained because they are unbounded in size and energy
and are not required to satisfy any radial boundary conditions.  It is
therefore uncertain whether they could be realized as intermediate
asymptotic forms in practice.

In a recent paper, Blandford \& Begelman (1999) have sought to resolve
both these difficulties by elaborating on the basic model, adding
terms that parametrize the removal of mass, angular momentum and
energy by an outflow, which is considered necessary `to allow
accretion to proceed'.  In their solution the flow is rotating even
though $\gamma=5/3$, and the Bernoulli parameter can be negative.
However, this was again achieved by adjusting the scalings so that
essentially no mass reaches the central object.

The purpose of this letter is to investigate the fluid dynamics of
quasi-spherical accretion from an alternative viewpoint which avoids
many of the strictures of steady self-similar solutions.  In the
theory of thin discs, a consideration of the initial-value problem, in
which a cool and narrow ring of viscous fluid orbiting around a
central mass spreads radially and is accreted, provides an excellent
demonstration of the dynamics of disc accretion and is in some ways
more informative than steady solutions.  In that problem the surface
density satisfies a diffusion equation which, depending on the
properties of the viscosity, is either linear, in which case the Green
function can be obtained analytically (L\"ust 1952; Lynden-Bell \&
Pringle 1974), or non-linear, in which case similarity methods can be
used to obtain an exact special solution of the equation (Pringle
1974; Lin \& Bodenheimer 1982; Lin \& Pringle 1987; Lyubarskii \&
Shakura 1987).  This self-similar solution accurately describes the
behaviour of all solutions of the initial-value problem long after the
inner radius of the fluid first reaches the central object.  In the
case of quasi-spherical accretion, the governing equations are
non-linear and similarity methods can again be used to provide an
exact solution which may similarly be expected to be an attractor for
the initial-value problem.

\section{Basic equations}

A simplified one-dimensional model for non-relativistic
quasi-spherical accretion is adopted in which physical quantities
depend only on the spherical radius $r$ and on time $t$.  The model is
described by the equations
\begin{equation}
{{{\rm D}\rho}\over{{\rm
D}t}}=-{{\rho}\over{r^2}}{{\partial}\over{\partial r}}(r^2u),\label{drho}
\end{equation}
\begin{equation}
\rho\left({{{\rm D}u}\over{{\rm
D}t}}-r\Omega^2\right)=-\rho{{{\rm d}\Phi}\over{{\rm d}r}}-{{\partial
p}\over{\partial r}},
\end{equation}
\begin{equation}
\rho{{{\rm D}\ell}\over{{\rm
D}t}}={{1}\over{r^2}}{{\partial}\over{\partial r}}\left(\mu
r^4{{\partial\Omega}\over{\partial r}}\right)
\end{equation}
and
\begin{equation}
\rho{{{\rm D}e}\over{{\rm
D}t}}=-{{p}\over{r^2}}{{\partial}\over{\partial r}}(r^2u)+\mu
r^2\left({{\partial\Omega}\over{\partial r}}\right)^2.\label{de}
\end{equation}
Here $\rho$ is the density, $u$ the radial velocity, $\Omega$ the
angular velocity, $\Phi=-GM/r$ the gravitational potential (where $M$
is the mass of the central object), $p$ the pressure, $\ell=r^2\Omega$
the specific angular momentum, $\mu$ the (dynamic) viscosity (which,
for simplicity, is assumed to act only on the differential rotation)
and $e$ the specific internal energy.  The Lagrangian time derivative
is
\begin{equation}
{{\rm D}\over{{\rm D}t}}={{\partial}\over{\partial
t}}+u{{\partial}\over{\partial r}}.
\end{equation}
It is further assumed that the fluid is an ideal gas
with
\begin{equation}
p=(\gamma-1)\rho e,\qquad1<\gamma\le5/3,
\end{equation}
and that the viscosity is given, as in NY, by an $\alpha$-model,
\begin{equation}
\mu=\alpha p/\Omega_{\rm K},
\end{equation}
where $\Omega_{\rm K}=(GM/r^3)^{1/2}$ is the Keplerian angular
velocity.  The parameters $\gamma$ and $\alpha$ are assumed to be
constant.

These equations may be obtained from the full axisymmetric equations
in spherical polar coordinates $(r,\theta,\phi)$ by considering the
equatorial plane $\theta=\pi/2$ and neglecting terms associated with
the $\theta$-velocity and any $\theta$-dependence.  The variables may
be considered as approximate spherically averaged quantities.
Although approximate, the equations have two important properties:
they reduce exactly to those of spherical accretion in the absence of
rotation, and they possess exact conservation laws for mass, angular
momentum and energy, which are of the form
\begin{equation}
{{\partial\rho}\over{\partial
t}}+{{1}\over{r^2}}{{\partial}\over{\partial r}}(\rho r^2u)=0,
\end{equation}
\begin{equation}
{{\partial}\over{\partial
t}}(\rho\ell)+{{1}\over{r^2}}{{\partial}\over{\partial r}}\left(
\rho\ell r^2u-\mu r^4{{\partial\Omega}\over{\partial
r}}\right)=0
\end{equation}
and
\begin{equation}
{{\partial}\over{\partial
t}}(\rho\varepsilon)+{{1}\over{r^2}}{{\partial
F_\varepsilon}\over{\partial r}}=0,
\end{equation}
where
\begin{equation}
\varepsilon={\textstyle{{1}\over{2}}}u^2+{\textstyle{{1}\over{2}}}
(r\Omega)^2+e+\Phi
\end{equation}
is the specific total energy and
\begin{equation}
F_\varepsilon=\rho\varepsilon r^2u+pr^2u-\mu
r^4\Omega{{\partial\Omega}
\over{\partial r}}
\end{equation}
the total energy flux.  These equations are appropriate to
quasi-spherical accretion, rather than disc accretion, because the
dissipated energy is retained in the fluid and may be advected with
the flow.

\section{Self-similar solution}

\subsection{Analysis}

NY solved essentially equations (\ref{drho})--(\ref{de}) for the case
of a steady, radially self-similar flow.  Such a solution is unbounded
in size, mass, angular momentum and energy, and must therefore be
interpreted as an approximation to a finite steady solution within a
region far removed from its boundaries.  In contrast, let us seek a
time-dependent solution of finite size, for which the integrals
representing the total mass, angular momentum and energy,
\begin{eqnarray}
M&=&4\pi\int_0^\infty\rho\,r^2\,{\rm d}r,\\
J&=&4\pi\int_0^\infty\rho\ell\,r^2\,{\rm d}r,\\
E&=&4\pi\int_0^\infty\rho\varepsilon\,r^2\,{\rm d}r,
\end{eqnarray}
are all convergent.  The central object is considered to be
arbitrarily small and to act as a sink for mass and (negative) energy
but not as a source for angular momentum.  Therefore $J$ is strictly
conserved, but $M$ and $E$ are not.

The time-dependent self-similar solution is found by a well-known
procedure (e.g. Barenblatt 1979).  In the standard terminology the
solution is of type I, which means that the similarity variable can be
deduced simply from dimensional considerations.  In this case the
dimensional constants present are the quantity $GM$ and the total
angular momentum $J$.  One therefore identifies the similarity
variable
\begin{equation}
\xi=r(GMt^2)^{-1/3}
\end{equation}
and seeks a solution of the form
\begin{eqnarray}
\rho(r,t)&=&\rho_*(\xi)\,J(GM)^{-5/3}t^{-7/3},\\
u(r,t)&=&u_*(\xi)\,(GM)^{1/3}t^{-1/3},\\
\Omega(r,t)&=&\Omega_*(\xi)\,t^{-1},\\
e(r,t)&=&e_*(\xi)\,(GM)^{2/3}t^{-2/3}.
\end{eqnarray}
In such a solution each physical quantity retains a similar spatial
form as the flow evolves, but the characteristic length-scale of the
flow increases proportionally to $t^{2/3}$.  Note that the total mass
$M$ is proportional to $t^{-1/3}$ and the total energy $E$ (which is
in fact negative) proportional to $t^{-1}$.

Substitution into equations (\ref{drho})--(\ref{de}) yields the
dimensionless equations
\begin{equation}
(u_*-{\textstyle{{2}\over{3}}}\xi)\rho_*^\prime-{\textstyle{{7}\over{3}}}
\rho_*=-\rho_*\xi^{-2}(\xi^2u_*)^\prime,
\end{equation}
\begin{equation}
\rho_*\left[(u_*-{\textstyle{{2}\over{3}}}\xi)u_*^\prime-
{\textstyle{{1}\over{3}}}u_*-\xi\Omega_*^2+\xi^{-2}\right]=
-(\gamma-1)(\rho_*e_*)^\prime,
\end{equation}
\begin{eqnarray}
\lefteqn{\rho_*\left[(u_*-{\textstyle{{2}\over{3}}}\xi)(\xi^2\Omega_*)
^\prime+{\textstyle{{1}\over{3}}}\xi^2\Omega_*\right]=}&\nonumber\\
&&(\gamma-1)\alpha\xi^{-2}(\rho_*e_*\xi^{11/2}\Omega_*^\prime)^\prime
\end{eqnarray}
and
\begin{eqnarray}
\lefteqn{(u_*-{\textstyle{{2}\over{3}}}\xi)e_*^\prime-
{\textstyle{{2}\over{3}}}e_*=}&\nonumber\\
&&-(\gamma-1)e_*\xi^{-2}(\xi^2u_*)^\prime+(\gamma-1)\alpha
e_*\xi^{7/2}(\Omega_*^\prime)^2,
\end{eqnarray}
where a prime denotes differentiation with respect to $\xi$.  The
constraint
\begin{equation}
4\pi\int_0^\infty\rho_*\xi^2\Omega_*\,\xi^2\,{\rm d}\xi=1
\end{equation}
provides a normalization condition for the density.

This is a fifth-order system of non-linear ordinary differential
equations.  Critical points occur wherever
\begin{equation}
u_*-{\textstyle{{2}\over{3}}}\xi=0\quad\hbox{or}\quad\pm[\gamma
(\gamma-1)e_*]^{1/2},
\end{equation}
i.e. wherever the radial velocity measured with respect to a
self-similarly expanding coordinate system is equal to zero or to the
local sound speed.

\subsection{Inner limit}

When $\gamma<5/3$, an appropriate asymptotic solution as $\xi\to0$ is
of the form
\begin{eqnarray}
\rho_*(\xi)&\sim&\xi^{-3/2}(A_\rho+B_\rho\xi^\lambda+\cdots),\\
u_*(\xi)&\sim&\xi^{-1/2}(A_u+B_u\xi^\lambda+\cdots),\\
\Omega_*(\xi)&\sim&\xi^{-3/2}(A_\Omega+B_\Omega\xi^\lambda+\cdots),\\
e_*(\xi)&\sim&\xi^{-1}(A_e+B_e\xi^\lambda+\cdots),
\end{eqnarray}
in which
\begin{eqnarray}
A_u&=&-{{(9\gamma-5)}\over{9(\gamma-1)}}{{g}\over{\alpha}},\\
A_\Omega^2&=&{{2(5-3\gamma)(9\gamma-5)g}\over{81(\gamma-1)^2\alpha^2}},\\
A_e&=&{{2(9\gamma-5)g}\over{27(\gamma-1)^2\alpha^2}},
\end{eqnarray}
where
\begin{equation}
g=\left[1+{{162(\gamma-1)^2\alpha^2}\over{(9\gamma-5)^2}}\right]^{1/2}-1,
\end{equation}
while $A_\rho$ is determined subsequently from the density
normalization.  The parameter $\lambda$ (such that $0<\lambda\le1$)
and the vector $[B_\rho,B_u,B_\Omega,B_e]^{\rm T}$ satisfy a certain
algebraic eigenvalue problem.  At leading order, we have essentially
the solution of NY.  This is reasonable because, for small $\xi$, the
flow has experienced many orbits and, except for the declining
density, may be expected to approach a steady state.

When $\gamma=5/3$, we have instead
\begin{eqnarray}
\rho_*(\xi)&\sim&\xi^{-3/2}(\tilde A_\rho+\tilde B_\rho\xi+\cdots),
\label{rhoinner2}\\
u_*(\xi)&\sim&\xi^{-1/2}(\tilde A_u+\tilde B_u\xi+\cdots),\\
\Omega_*(\xi)&\sim&\xi^{-1/2}(\tilde B_\Omega+\cdots),\\
e_*(\xi)&\sim&\xi^{-1}(\tilde A_e+\tilde B_e\xi+\cdots),\label{einner2}
\end{eqnarray}
in which
\begin{equation}
\tilde A_u=-{{5\tilde g}\over{\alpha}}\qquad\hbox{and}\qquad
\tilde A_e={{15\tilde g}\over{\alpha^2}},
\end{equation}
where
\begin{equation}
\tilde g=\left(1+{{2\alpha^2}\over{25}}\right)^{1/2}-1.
\end{equation}

The limit $\gamma\to5/3$ is a singular one because $\lambda\to0$ so
that $\tilde A_u\ne\lim A_u$ and $\tilde A_e\ne\lim A_e$.  In other
words, the NY solution is subject to fractional corrections which
become of order unity at all radii as $\gamma\to5/3$, and the Bondi
solution is never attained.  The further development of the inner
solution consists in general of an irregular power series which is
beyond the scope of this paper.

\subsection{Outer limit}

Any solution of finite size must possess a free surface at some point
$\xi=\xi_{\rm s}$.  At such a point the velocity must equal $u=({\rm
d}r/{\rm d}t)_\xi$, which implies
$u_*={\textstyle{{2}\over{3}}}\xi_{\rm s}$.  Assuming that the entropy
tends to a finite limit, the surface resembles that of a polytrope.
In terms of $s=\xi_{\rm s}-\xi$, we have
\begin{eqnarray}
\rho_*(\xi)&\sim&s^{1/(\gamma-1)}(C_\rho+D_\rho s+\cdots),\\
u_*(\xi)&\sim&{\textstyle{{2}\over{3}}}\xi_{\rm s}+D_u s+\cdots,\\
\Omega_*(\xi)&\sim&C_\Omega+D_\Omega s+\cdots,\\
e_*(\xi)&\sim&s(C_e+D_e s+\cdots).
\end{eqnarray}
All the coefficients in this expansion, except for the density
normalization, follow algebraically from a knowledge of $\xi_{\rm s}$.
The expansion fails if $\gamma<9/7$.

\subsection{Numerical solution}

\begin{figure*}
\centerline{\epsfbox{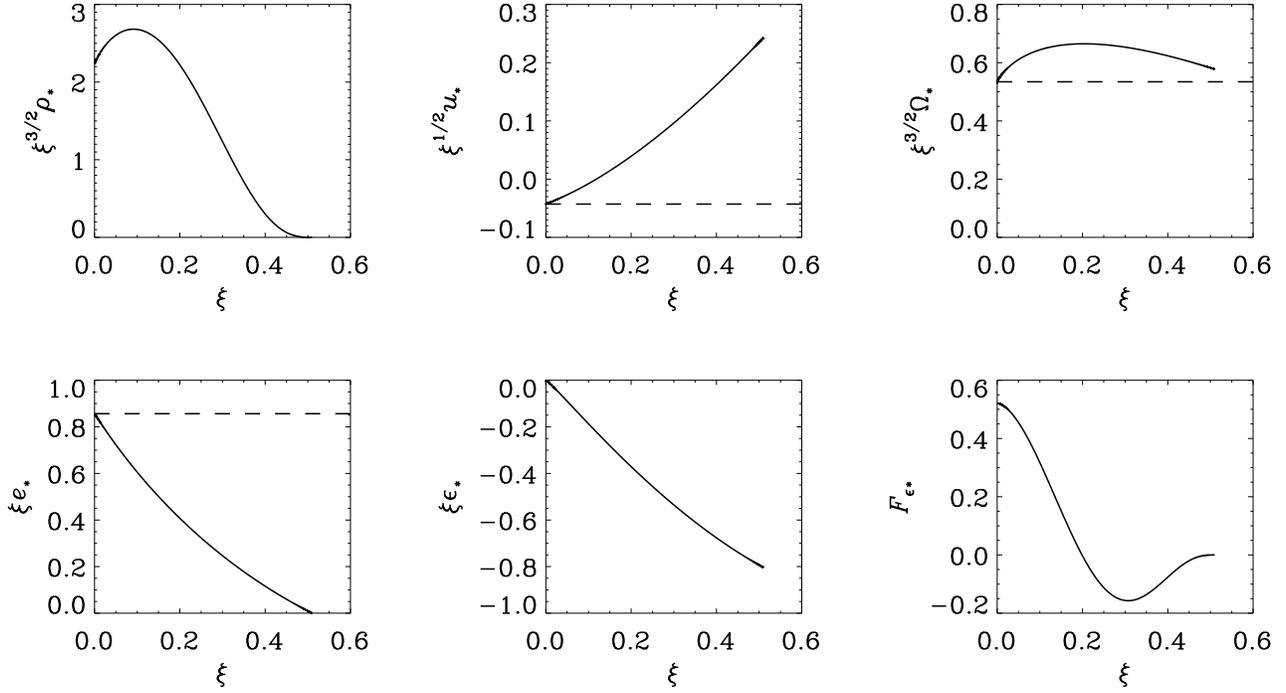}}
\caption{Time-dependent self-similar solution for $\gamma=4/3$ and
$\alpha=0.1$.  The surface occurs at $\xi_{\rm s}\approx0.5101$.  The
quantities $\varepsilon_*$ and $F_{\varepsilon*}$, defined by
$\varepsilon=\varepsilon_*(GM)^{2/3}t^{-2/3}$ and
$F_\varepsilon=F_{\varepsilon*}Jt^{-2}$, are dimensionless versions of
the specific total energy and the total energy flux.  Note that
$\xi^{3/2}\Omega_*$ is the ratio of the angular velocity to the
Keplerian angular velocity, while $\xi e_*$ is proportional to the
ratio of the temperature to the virial temperature.  The dashed lines
represent the steady solution of NY, which is the limiting form of the
present solution as $\xi\to0$.}
\end{figure*}

\begin{figure*}
\centerline{\epsfbox{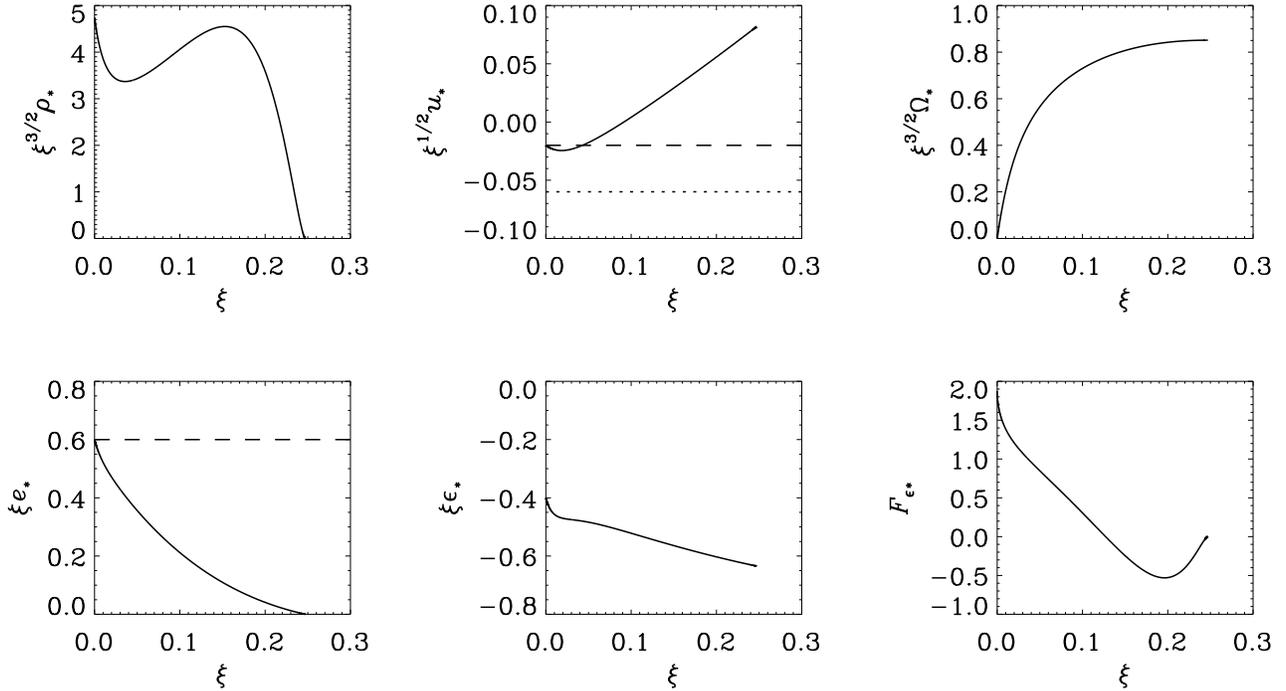}}
\caption{Time-dependent self-similar solution for $\gamma=5/3$ and
$\alpha=0.1$.  The surface occurs at $\xi_{\rm s}\approx0.2470$.  The
dashed lines represent the limiting form
(\ref{rhoinner2})--(\ref{einner2}) of the solution.  The dotted line
represents the radial velocity according to NY.}
\end{figure*}

If the value of $\xi_{\rm s}$ is guessed, the equations can be
integrated inwards from a point very near the surface, using the above
expansion.  By adjusting $\xi_{\rm s}$ the solution that has the
correct asymptotic behaviour as $\xi\to0$ can be identified.  By this
method satisfactory solutions have been obtained without internal
critical points.

Examples of these solutions are shown in Figs~1 and~2.  There are
several points of interest.  For $\gamma<5/3$, the solution matches
only asymptotically on to that of NY, deviating substantially from it
at larger radii.  In the case $\gamma=5/3$ the flow is differentially
rotating, although it is highly sub-Keplerian at small radii.  For
$\alpha\ll1$, as in these examples, the radial velocity is everywhere
highly subsonic with respect to a self-similarly expanding coordinate
system.  Although the temperature is comparable to the virial
temperature over much of the flow, this does not drive a supersonic
outflow at large radii.  Finally, the flows are bound to the central
object in the sense that the total energy is negative.  As in a thin
disc, there is a net outflow of energy from the centre of the disc
because the central object is accreting bound matter.

\section{Discussion}

In this paper the equations of time-dependent quasi-spherical
accretion have been solved by semi-analytical similarity methods.  The
solution describes asymptotically the fate of an initially cool and
narrow ring of viscous fluid orbiting around a central mass when it is
unable to radiate efficiently.  The flow is finite in size, mass,
energy and angular momentum and avoids many of the strictures of the
steady self-similar solution of NY.

In particular, a solution was found for the important case
$\gamma=5/3$ that has differential rotation and viscous dissipation,
and resembles a disc at large radii.  No barrier to accretion was
found to exist in this case.  At a fixed radius, the flow gradually
becomes increasingly sub-Keplerian (Fig.~3).  The limit $\gamma\to5/3$
was shown to be singular so that the Bondi solution is not attained.
This conclusion applies also to steady solutions that are not
self-similar.

\begin{figure}
\centerline{\epsfbox{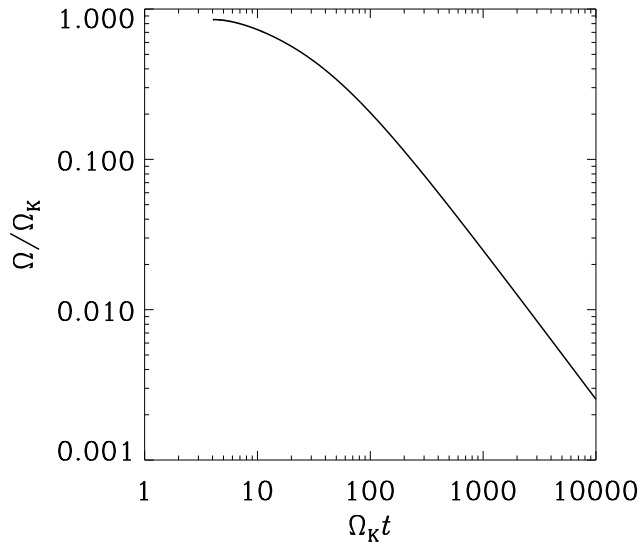}}
\caption{The variation with time of the ratio of the angular velocity
to the Keplerian angular velocity at any fixed radius, for the same
solution as in Fig.~2.}
\end{figure}

The flows are bound to the central object in the sense that the total
energy is negative.  Indeed, for $\gamma\ge4/3$ the specific total
energy is negative at all radii.  It is suggested that previous
interpretations based on the sign of the Bernoulli parameter may be
insecure because Bernoulli's theorem does not hold for a viscous fluid
(nor for a time-dependent one, as here).  The meaning of the Bernoulli
parameter is especially unclear in a flow that is driven specifically
by viscous stresses.

The present model admits, at least in principle, the possibility of a
supersonic outflow at large radii, but this was not found to be
necessary.  This suggests, contrary to Blandford \& Begelman (1999),
that quasi-spherical accretion can take place without the need for
powerful outflows.  However, because the model is topologically
restricted, this conclusion must be regarded as tentative and one
cannot exclude the possibility that an outflow would nevertheless
occur from the hot inner part of the flow if the one-dimensional
assumption were relaxed.  This might be answered by restoring the
latitudinal dependence in the equations and finding whether the
analogous time-dependent self-similar solution includes a bipolar
outflow.  However, this would require the solution of a non-linear
free-boundary problem of elliptic (or possibly mixed) type, which is
technically demanding.  It is perhaps more likely that the
initial-value problem with latitudinal dependence could be solved
using a time-dependent numerical method.  Indeed, Igumenshchev, Chen
\& Abramowicz (1996) have already performed a calculation of this
type, adopting $\gamma=4/3$ and starting from a thick torus.  They
found rather complex behaviour, including convection, but no evidence
for bipolar outflows.

The time-dependence of the flow is not essential to the conclusions of
this paper.  However, the present method of investigation allows a
description of generic properties of solutions of the initial-value
problem that is essentially free from infinities and contains no
undetermined parameters, only the physical ones $\gamma$ and $\alpha$.
Questions concerning the boundness of the flow and the necessity of
outflows can be addressed with greater confidence within such a
framework.

\section*{Acknowledgments}

I am grateful to Henk Spruit for suggesting this problem and for
helpful discussions.  This work was supported by the Training and
Mobility of Researchers Network `Accretion on to Black Holes, Compact
Stars and Protostars'.

\label{lastpage}

\end{document}